# Young Stellar Clusters and Star Formation Throughout the Galaxy

**A Science Working Paper for the Astro2010 Survey**
**Planetary Systems and Star Formation and**
**The Galactic Neighborhood Science Panels**

**Eric Feigelson*[1], Fred Adams[2], Lori Allen[3], Edwin Bergin[2], John Bally[4], Zoltan Balog[5], Tyler Bourke[6], Crystal Brogan[7], You-Hua Chu[8], Edward Churchwell[9], Marc Gagne[10], Konstantin Getman[1], Todd Hunter[7], Larry Morgan[7], Philip Massey[11], Mordecai-Mark Mac Low[12], Eric Mamajek[13], S. Thomas Megeath[14], C. Robert O'Dell[15], Jill Rathborne[6], Luisa Rebull[16], Steven Stahler[17], Leisa Townsley[1], Junfeng Wang[6], Jonathan Williams[18]**

[1] Penn State University [2] University of Michigan [3] National Optical Astronomy Observatory
[4] University of Colorado [5] University of Arizona [6] Harvard-Smithsonian Center for Astrophysics
[7] National Radio Astronomy Observatory [8] University of Illinois [9] University of Wisconsin
[10] West Chester University [11] Lowell Observatory [12] American Museum of Natural History
[13] University of Rochester [14] University of Toledo [15] Vanderbilt University
[16] Spitzer Science Center [17] University of California Berkeley [18] University of Hawaii

* Contact person: Eric Feigelson, Department of Astronomy & Astrophysics, Penn State University. Email: edf@astro.psu.edu 814-865-0162

## Introduction

Most stars are born in rich young stellar clusters (YSCs) embedded in giant molecular clouds (Lada & Lada 2003). They represent the dominant mode of star formation after the primordial era. The most massive stars live out their short lives there, profoundly influencing their natal environments by ionizing HII regions, inflating wind-blown bubbles and champagne flows, and soon exploding as supernovae (Beuther et al. 2008). Their ejecta chemically enrich the interstellar medium (ISM), drive turbulence, power the Galaxy's cosmic ray flux, trigger secondary cluster formation, and carve superbubbles extending into the Galactic halo (Cesaroni et al. 2005). On large scales, associations of massive stars produce supernovae that power super-bubbles which propel the energetics of the disk and halo interstellar medium. Converging flows where superbubbles intersect promote molecular cloud formation and the next generation of star formation. In starburst environments, super-star clusters drive nuclear super-winds that heat, ionize and chemically enrich the intergalactic medium.

On smaller scales, the expanding HII regions and supernova remnants from massive stars paradoxically trigger new star formation as they destroy their natal clouds (Elmegreen & Palous 2007). Thousands of lower-mass pre-main sequence stars accompany the massive stars in each YSC, both in the dense clusters around OB stars and in the surrounding molecular cloud. Planet formation may be qualitatively different in YSC environments compared to isolated star formation environments, but it is not clear whether planet growth is inhibited or promoted in the vicinity of OB stars (Hollenbach et al. 2000; Thorpe & Bally 2004). There is evidence -- based on the truncated Kuiper Belt, H-poor Uranus and Neptune, and short-lived radionuclides -- that our Solar System formed in such a rich cluster (Hester et al. 2004).

While this schematic picture of YSCs is established, our understanding of the astrophysical processes at work is still quite primitive. Do clusters form rapidly during a single collapse event or slowly over many crossing times (Elmegreen 2000; Tan et al. 2006)? Why are massive stars rare, and do they via accretion disks or stellar mergers (Zinnecker & Yorke 2007)? How does the feedback from OB stars both halt and promote further star formation? Is the stellar Initial Mass Function (IMF) truly universal over a wide range of cloud conditions, and what produces its distinctive shape (Bonnell et al. 2007)? When and why does primordial mass segregation (i.e. concentration of the OB stars into the core) occur (Huff & Stahler 2007)? What effect do shocked OB winds have on the physics of the HII region and the confining molecular cloud (Townsley et al. 2003)? What fraction of stars in the Galaxy form from triggered processes? What determines whether a young stellar cluster survives the dispersal of its parental molecular gas and becomes a bound open or globular cluster? How does the cluster environment influence the evolution of protoplanetary disks and subsequent formation of planetary systems?

While considerable progress was made in understanding the formation of *isolated* stars during the past two decades, the more complex phenomena involved in *clustered* star formation have only just begun to be elucidated. Powerful new datasets are now emerging from both space-based (Spitzer and Chandra) and ground-based (sub-millimeter mapping, wide-field infrared imaging and multi-object spectroscopy) surveys. The technologies are challenging: YSC observational studies require both high spatial resolution and wide fields at wavelengths that penetrate obscuring molecular material and remove contaminating Galactic field stars. New surveys are reaching beyond the closest clusters in Orion to study massive star forming regions along the Galaxy's nearby spiral arms and in Local Group galaxies, providing critical links to extragalactic HII regions and starburst galaxies. The proliferation of theories concerning, for example, OB star formation and the establishment of the IMF, can soon be tested.

The 2010-19 decade promises to witness breakthroughs in our understanding of star formation at this crucial intermediate scale between individual stars and entire galaxies. Studying young clusters, particularly those embedded in their natal clouds, requires multi-wavelength observations ranging from hard X-rays to radio waves. Young clusters are laboratories where the birth of stars and planets can be studied with large statistical samples, feedback mechanisms can be investigated, and foundations of cosmic starburst phenomena can be decoded. We outline here eight scientific issues ready for major advances in the coming decade.

**The IMFs and structures of rich stellar clusters**
The IMF in local YSCs and in older stellar populations appears to be a Salpeter powerlaw distribution at masses above ~1 $M_O$ appended to lognormal distribution peaking around 0.3 $M_O$ (Chabrier 2003). However, some of the most massive clusters in the Local Group appear to have a top-heavy IMF (Harayama et al. 2008). Understanding the conditions of OB star formation and the physical origin of the IMF impacts our understanding of starburst galaxies, the first stars, and the Epoch of Reionization. Critical clues will emerge from infrared and X-ray surveys that sample the lower mass populations (Figures 1-4); morphological studies can then examine smooth vs. clumpy structures and the presence of mass segregation. Kinematical investigations using precision astrometry and radial velocities will complement morphological studies to challenge theoretical models of turbulent cloud collapse and early cluster evolution.

**Triggered star formation around young stellar clusters**
In the past, it had been speculated that the impact of supernova shocks on molecular clouds were a major trigger of secondary star formation. While this now looks unlikely (Joung & Mac Low 2006), recent research has instead pointed to the importance of ionization and wind shocks from pre-supernova OB stars. Triggered populations are found both within small bright-rimmed clouds and larger cloud structures on the edges of expanding HII regions using infrared and X-ray images (Figure 3). The evidence supports theoretical `collect-and-collapse' models on large scales (Elmegreen & Lada 1977) and `radiative driven implosion' on small scales (Bertoldi 1989). Nonetheless, the quantitative importance of triggering in Galactic star formation has not been established and remains a future challenge.

**The fate of OB winds**
The inner regions of radiatively accelerated OB winds can be studied with ultraviolet P Cygni absorption lines and X-ray emission lines. But in their outer regions, the winds can only be observed indirectly. However, with sensitive high-resolution X-ray images, the faint emission from shocked OB winds can be seen on parsec scales (Townsley et al. 2003). The hot gas flowing from central cluster in M 17 (Figure 1) is a prime example of these X-ray flows and reveals the winds' confinement by and escape from the molecular cloud environment. The astrophysics of HII regions must be reconsidered to account for interiors filled with $10^7$ K rather than $10^4$ K plasma.

**The stellar populations of Infrared Dark Clouds (IRDCs)**
Studies of the Galactic Plane at mid-infrared wavelengths during the 1990s first identified IRDCs as dark extinction features seen in absorption against the bright mid-IR emission arising from the Galactic (Carey et al. 1998; see Figure 2). IRDCs are ubiquitous across the Galaxy with filamentary morphologies, very high densities ($n_{H2} > 10^5$ $cm^{-3}$, $N_H \sim 10^{23}$-$10^{25}$ $cm^{-2}$), and low temperatures (T<25 K) It is argued that IRDCs are short-lived cold gaseous precursors of high-mass stars and stellar clusters; they may be responsible for a significant fraction of the star formation in the Galaxy (Rathborne et al. 2006). Yet, due to their opacities and distances, their young stellar populations are poorly constrained beyond a handful of infrared-luminous protostars. Radio, mid-infrared and X-ray surveys will have the sensitivity to pierce through both intervening and local cloud material to detect the stellar populations of IRDCs at all stages of pre-main sequence evolution. The star formation efficiency, history, and IMF in IRDCs can then be compared to normal molecular clouds.

**The most massive star clusters in the Galaxy**
From the spatial distribution of known stellar clusters and comparison with other spiral galaxies, we know that only a small fraction of the ~20,000 YSCs in the Galactic disk have been identified to date (Figer 2008). In particular, only a dozen of the most massive clusters ($M > 10^4$–$10^5$ $M_O$) are identified, such as NGC 3603 and the Arches Cluster. These clusters harbor the most massive ($M \sim 100$ $M_O$) main sequence, supergiant and Wolf-Rayet stars with prodigious winds and supernova remnants that have the greatest influence on the galactic interstellar medium. Radio and millimeter studies pierce the material surrounding the earliest stages of massive star formation. Deep, high-spatial resolution mid-infrared and X-ray studies are needed to locate and characterize the spatial, mass and age distributions of stars in these clusters. Their relationship to local interstellar environments is also critical to test theories of their origins; e.g. whether top-heavy IMFs preferentially form where the interstellar pressure is unusually high (McKee & Tan 2002).

**Tracing star formation throughout the Galactic Disk**
Wide-field surveys at penetrating wavelengths have revealed a remarkable medley of interstellar structures that trace star cluster formation across the Galactic Plane. The first indication was 21-cm line studies of expanding supershells and worms in the plane (Heiles 1979), leading to a model of a thousand chimneys exhausting wind and supernova ejecta into the Galactic halo (Norman & Ikeuchi 1989). More recently, the GLIMPSE mid-infrared survey with the Spitzer Space Telescope and the MAGPIS centimeter-wavelength survey with the Very Large Array reveal hundreds of previously unknown wind-swept, ionized bubbles and supernova remnants (Figure 5). Each of these are likely produced by yet-unknown young stellar clusters. Most star formation occurs in the inner 5 kpc of the Galaxy, and many star formation regions lie on the uncharted far side of the Galactic disk.

**The Galactic Center region**
Were it viewed from afar, the Milky Way would be seen as a mild nuclear starburst galaxy. Many of the most massive star, massive stars, and giant molecular clouds reside in the inner 1–100 pc. The molecular clouds have internal turbulent motions 5 times greater than seen in nearby clouds (Miyazaki & Tsuboi 2000). The intercloud medium in this region is extraordinarily hot and dense from thousands of past supernova explosions with a pressure $\sim 10^3$ times that of the medium near the Sun (Muno et al. 2004). Though difficult to discern, a supernova-driven wind appears to be emerging perpendicular to the disk (Law et al. 2009). Most of the massive star formation in the region has yet to be identified due to the observational difficulties of heavy obscuration and crowding (Figure 6).

**Star clusters in the Magellanic Clouds**
Massive star clusters in the Large and Small Magellanic Clouds (MCs) are more accessible than distant Galactic clusters due to low intervening obscuration, but are ~10 times more distant. The high-mass star formation rate in the MCs are an order of magnitude higher than in the Milky Way Galaxy, and thus serve as an invaluable bridge to understanding cosmic starburst phenomena (Figure 7). Studies of the most extreme super-star clusters will illuminate the birth of globular clusters and low-metallicity star formation in the early Universe. With current technology, global properties of major clusters are well-established, individual OB/WR stars can be studied, and the more luminous low-mass pre-main sequence objects can be detected. 30 Dor and other active complex give a uniquely detailed view of the effects of multiple supernova explosions. The high sensitivity and sub-arcsecond resolution of JWST and 8m-class telescopes are essential to make major steps forward in MC research.

## Programmatic recommendations

YSC studies are central to our understanding of star and planet formation across the Galaxy and throughout cosmic history. The astrophysics of star formation and pre-main sequence evolution has proved to be far more complex than early concepts that involved only gravitational collapse and convective stellar interiors. While individual protoplanetary disks and exoplanetary systems are best studied close to the Sun, most stars likely experience hostile YSC environments with OB radiation and SN ejecta during a portion of the planet formation epoch. Starburst investigations, whether in nearby galaxies or at high redshifts, require understanding of local YSCs to derive astrophysical meaning from correlations between global quantities observable from great distances.

YSC studies are poised to make major progress during the 2010-20 decade, but will require a wide range of scientific capabilities. Some of these are already underway including JWST, EVLA, VLBA, ALMA, Herschel, SOFIA and Chandra. Here we outline five urgent needs which need policy support:

**Development of a ground-based 30m-class adaptive-optics infrared telescope** Perhaps the most critical instrument needed to propel our understanding of star formation in our Galaxy is a 30m-class telescope such as the GMT, TMT and EELT. Galactic Plane fields are extremely crowded and the <0.2" resolution of these telescopes are essential to resolve obscured and distant YSCs. Siting and instrumentation efforts should facilitate observations in the mid-infrared regime, as even the most luminous members of distant clusters are almost inaccessible in the JHK bands (Figure 6). Mid-IR imaging is also needed for protoplanetary disk studies.

**Support for high-resolution X-ray telescopes** X-ray images of YSCs provide rich samples of young stars with little contamination by Galactic field stars. They also give unique information on the hot winds and supernovae that feed back energy and metals into the star forming gas, enriching and energizing the ISM. Given the astronomical community's investiment in high spatial resolution imaging with JWST and ALMA, the Chandra X-ray Observatory should continue operations through the decade support these new projects. to extend existing photon-starved YSC studies, and to extend X-ray imaging to more distant and obscured YSCs. Strong support is needed for the International X-ray Observatory, and technology development for the Generation-X mission should proceed energetically; its planned 0.1” resolution and 500x-Chandra sensitivity is superbly suited to YSC studies. The proposed Wide Field X-ray Telescope will give a unique scan of stellar clusters in the Galactic Plane.

**Large-aperture sub-millimeter and far-infrared telescopes** Characterizing the molecular cloud precursors of YSCs requires submillimeter line and continuum mapping high resolution and wide-area coverage. These include a large filled-aperture telescope such as the 30m CCAT planned for Chile, the LMT in Mexico, and focal-plane array upgrades to ALMA and the GBT. A space-based far-IR telescope would extend capabilities into the critical 50-150 μm range.

**Development of multi-object infrared spectrographs** Although imaging and broad-band photometry are essential first steps in discovering YSCs clusters throughout the Galactic disk, spectroscopy of individual stars (including low mass members) is needed to measure individual masses, ages, accretion, and metallicity, and radial velocities (e.g. Furesz et al. 2008). Multi-object spectroscopy must move into the infrared band: FLAMINGOS-2/Gemini and MOIRCS/Subaru are already underway, and plans are being developed for IRMOS/TMT.

**Support for theoretical modeling** Major commitments of time on high-performance computers are needed to extend theoretical simulations of YSC formation to realistically massive clusters (Bate et al. 2009). Simulations of YSC formation are beginning to incorporate the complex astrophysics of turbulent, partly-ionized magnetized molecular clouds, line and continuum radiative transfer, N-body gravitational interactions, and OB star radiation pressure, ionization and wind feedback. More analytical theory is also critically needed. Funding for individual investigator programs, supporting both theory and observational work at ground-based telescopes, should be strengthened.

Alves & Homeier 2003, ApJ 589, L45
Bate et al. 2003, MNRAS 392, 590
Bertoldi 1989, ApJ 346, 735
Beuther et al. (eds) 2008, Massive Star Formation
Bonnell et al. 2007, in Protostars & Planets V, 149
Briceno et al. 2007, in Protostars & Planets V, 345
Carey et al. 1998, ApJ 508, 721
Cesaroni et al. 2007, in Protostars & Planets V, 197
Chabrier 2003, PASP 115, 763
Churchwell et al. 2006, ApJ 649, 759
Deharveng, Lefloch et al. 2006, A&A 458, 191
Elmegreen & Lada 1977, ApJ 214, 725
Elmegreen 2000, ApJ 530, 277
Elmegreen & Palous (eds) 2007, IAU Symp 237
Feigelson et al. 2007, in Protostars & Planets V, 313
Figer 2008, in IAU Symp 250, 247
Furesz, Hartmann et al. 2008, ApJ 676, 1109
Harayama et al. 2008, ApJ 675, 1319
Heiles 1979, ApJ 229 533
Helfand, Becker et al. 2006, AJ 131, 2525
Hester et al. 2004, Science 304, 111
Hollenbach et al. 2000, in PIV, 401
Huff & Stahler 2007, ApJ 666, 281
Joung & Mac Low 2006, ApJ 653, 1266
Koenig, Allen et al. 2008, ApJ 688, 1142
Lada & Lada 2003, ARAA 41, 57
Law, Backer et al. 2009, arXiv:0901.1480
McKee & Tan 2002, Nature 416, 59
Miyazaki & Tsuboi 2000, ApJ 536, 357
Muno, Baganoff et al. 2004, ApJ 613, 326
Norman & Ikeuchi 1989, ApJ 345, 372
Rathborne, Jackson et al. 2005, ApJ 630, L181
Rathborne, Jackson et al. 2006, ApJ 641, 389
Tan, Krumholz & McKee 2006, ApJ 641, L121
Throop & Bally 2005, ApJ 623, L149
Townsley, Feigelson et al. 2003, ApJ 593, 874
Zinnecker & Yorke 2007, ARA&A 45, 481